\documentclass[aps,prb,twocolumn,superscriptaddress,amsmath,amssymb,reprint]{revtex4-2}

\usepackage[pdftex,colorlinks=true,hypertexnames=false,citecolor=blue]{hyperref}
\usepackage{graphicx,color,epsfig,rotate}
\usepackage{units}
\usepackage{float}
\usepackage{dcolumn}							
\usepackage{multirow}
\usepackage{xspace}								
\hypersetup{pdfpagemode=UseNone,pdfpagelayout=OneColumn,pdfstartview=FitH}

\newcolumntype{d}[1]{D{.}{.}{#1}} 

\newcommand{\TN}[0]{\ensuremath{T_N}\xspace}			
\newcommand{\muB}[0]{\ensuremath{\mu_{\text{B}}}\xspace}		

\newcommand{\SL}[0]{\ensuremath{1/T_1}\xspace}					

\newcommand{\Hparc}[0]{\ensuremath{H||c}\xspace}
\newcommand{\Hperpc}[0]{\ensuremath{H{\perp}c}\xspace}

\newcommand{\NaYbSe}[0]{NaYbSe$_{2}$\xspace}

\newcommand{\Na}[0]{$^{23}$Na\xspace}
\newcommand{\Se}[0]{$^{77}$Se\xspace}
\newcommand{\Cu}[0]{$^{63}$Cu\xspace}
\newcommand{\Al}[0]{$^{27}$Al\xspace}

\newcommand{\NaYbSeT}[0]{\texorpdfstring{NaYbSe$_2$}{NaYbSe2}\xspace}	
\newcommand{\NaT}[0]{\texorpdfstring{$^{23}$Na}{23Na}\xspace}			
\newcommand{\SeT}[0]{\texorpdfstring{$^{77}$Se}{77Se}\xspace}			

\begin{document}

\title{Anisotropic magnetism and spin fluctuations in the triangular-lattice spin-liquid candidate \NaYbSeT : a single-crystal \NaT and \SeT NMR study}

\author{S. Luther}
\email{s.luther@hzdr.de}
\affiliation{Hochfeld-Magnetlabor Dresden (HLD-EMFL) and W\"urzburg-Dresden
Cluster of Excellence ct.qmat, Helmholtz-Zentrum Dresden-Rossendorf, 01328 Dresden, Germany}

\author{K. M. Ranjith}
\affiliation{Max Planck Institute for Chemical Physics of Solids, 01187 Dresden, Germany}

\author{Th. Doert}
\affiliation{Faculty of Chemistry and Food Chemistry, TU Dresden, 01062 Dresden, Germany}

\author{J. Wosnitza}
\affiliation{Hochfeld-Magnetlabor Dresden (HLD-EMFL) and W\"urzburg-Dresden
Cluster of Excellence ct.qmat, Helmholtz-Zentrum Dresden-Rossendorf, 01328 Dresden, Germany}
\affiliation{Institut f\"ur Festk\"orper- und Materialphysik, TU Dresden, 01062 Dresden, Germany}

\author{H. K\"uhne}
\affiliation{Hochfeld-Magnetlabor Dresden (HLD-EMFL) and W\"urzburg-Dresden
Cluster of Excellence ct.qmat, Helmholtz-Zentrum Dresden-Rossendorf, 01328 Dresden, Germany}

\author{M. Baenitz}
\affiliation{Max Planck Institute for Chemical Physics of Solids, 01187 Dresden, Germany}

\date{\today}

\begin{abstract}
The ytterbium-based delafossite \NaYbSe is discussed as a prototype for a spin-orbit entangled, effective spin-1/2 triangular spin lattice with emerging
antiferromagnetic correlations and a quantum-spin-liquid (QSL) ground state. We report on a comprehensive study of the static and dynamic anisotropic magnetism in single-crystalline samples of \NaYbSe, using NMR spectroscopy as a local-probe technique. We performed \Na and \Se NMR measurements in 
magnetic fields up to 16 T, applied along the in-plane and out-of-plane crystallographic directions and at temperatures from 300 down to 0.3 K. We could determine the anisotropic hyperfine contributions from the angular
dependence of the \Na and \Se NMR spectra. In the paramagnetic regime, we probed the temperature dependence 
of the \Na and \Se spectral shift and the hyperfine coupling constants for fields applied along the principal crystal axes. The
spin-lattice relaxation-rate data indicate critical spin fluctuations and the absence of long-range magnetic order at low magnetic fields and temperatures down to 0.3 K, evidenced by a monotonic increase of \SL and associated spectral broadening. This is a clear proof of the evolution of a critical QSL 
ground state with residual fluctuations down to lowest temperatures. At elevated fields, we observe the emergence of long-range order, as the temperature-dependent \SL rate passes through a pronounced maximum at \TN at given field, followed by a decrease at lower temperatures. Further, we find an inhomogeneous broadening of the \Na spectra below \TN, probing the histogram of the local-field distribution in the presence of the field-induced order.
\end{abstract}

\maketitle

\section{Introduction}

In recent years, there has been a marked increase in interest in quantum magnets with frustrated spin lattices. These materials exhibit novel phenomena, such as quantum spin liquids (QSLs), which represent highly entangled ground states characterized by significant quantum fluctuations and the absence of magnetic order at temperatures approaching absolute zero
\cite{Anderson_RVB, Balents_QSL, Savary_Balents_review, Zhou_QSL}. Triangular-lattice antiferromagnets with spin 1/2 are particularly promising candidates for hosting QSL ground states \cite{Shimizu_kappa_QSL, Itou_Et_QSL, Zhou_BaCuSbO_QSL}. Over the past years, ytterbium (Yb)-based triangular-lattice materials with antiferromagnetic interactions have been intensively investigated as QSL candidates.
There have been numerous experimental studies on the ground-state nature of YbMgGaO$_4$, claiming a gapless QSL state
\cite{Li_YbMgGaO4_2015_PRL, Li_YbMgGaO4_2015_SciRep, Li_YbMgGaO4_Muon_2016_PRL, Xu_YbMgGaO4_ThermCond_2016_PRL, Li_YbMgGaO4_CEF_2017_PRL, 
Li_YbMgGaO4_RVB_2017_Nature, Li_YbMgGaO4_ValenceBond_2019_PRL, Majumder_YbMgGaO4_PRR_2020, Ding_YbMgGaO4_muSR_PRB_2020, Rao_YbMgGaO4_2021_Nature, 
Shen_YbMgGaO4_spinonsurface_2016_Nature, Paddison_YbMgGaO4_excitations_2017_Nature, Shen_YbMgGaO4_spinonfield_2018_Nature}. However, a charge disorder
of the randomly distributed nonmagnetic Mg$^{2+}$ and Ga$^{3+}$ sites is discussed to have an influence on the ground state.
Consequently, there is growing interest in exploring other Yb-based materials that lack structural disorder, such as Yb(BaBO$_3$)$_3$ \cite{Yb(BaBO3)3_PRB} and several members of the delafossite family
\cite{Liu_rare_earth, Ranjith_NaYbO2, Bordelon_NaYbO2_neutron, Bordelon_NaYbO2_spin_excitation, Ding_NaYbO2, Ranjith_NaYbSe2, Dai_Spinon_NaYbSe2, 
Baenitz_NaYbS2, Sarkar_NaYbS2, Xing_CsYbSe2}. These delafossites feature a trigonal crystal structure with the space group $R\bar{3}m$ 
and a planar triangular arrangement of the Yb$^{3+}$ ions. A pronounced magnetic anisotropy in arises from strong spin-orbit coupling and crystalline-electric-field effects, resulting in the low-temperature formation of a well-separated Kramers ground-state doublet with an effective spin 1/2.

Triangular spin lattices with isotropic Heisenberg interactions are known to yield a ground state that exhibits 120$^{\circ}$ order. For applied magnetic fields, a phase diagram with different ordered spin configurations was proposed. These configurations include the canonical 120$^{\circ}$ state, the up-up-down (uud) state, and the canted uud arrangement \cite{Kawamura_phase, Seabra_phase}. The exchange interactions of Yb-based delafossites can be 
described by a pseudospin exchange model, which incorporates an XXZ Hamiltonian with bond-dependent interactions \cite{exchange_model_Schmidt}.
By considering these mechanisms and also next-nearest-neighbor Heisenberg interactions, it is possible to perturb the zero-field ordered ground state, suppress magnetic order, and induce a transition into the QSL state \cite{Zhu_Topography_QSL, Kaneko_phase}.

In a previous work, we determined the anisotropic, belly-shaped $H-T$ phase diagram of \NaYbSe by means of specific-heat and magnetization measurements \cite{Ranjith_NaYbSe2}. Furthermore, we deduced a spin anisotropy from susceptibility and ESR data at low temperatures. At low fields, the absence of magnetic order and persistent strong spin 
fluctuations, as well as a spinon Fermi surface were reported \cite{Zhang_muSR_NaYbSe2, Dai_Spinon_NaYbSe2}, suggesting a QSL ground state.
More recently, the coexistence of quasi-static and dynamic spins for out-of-plane fields has been observed \cite{Zhu_muSR_NMR_NaYbSe2}.

In the present manuscript, we report the results of our \Na and \Se NMR measurements of  \NaYbSe for a wide range of temperatures down to 0.3 K and magnetic fields up to 16 T, applied both in-plane and out-of-plane. We probe the presence of a critical QSL ground state with strong residual fluctuations by means of \SL data, showing by a monotonic increase and associated spectral broadening upon approaching low temperatures and low magnetic fields. At elevated fields, we observe the emergence of field-induced long-range order, manifested as a pronounced maximum at \TN at given field. The inhomogeneous local-field distribution, arising from the field-induced order, is probed via \Na spectroscopy.

\section{Experimental}

High-quality single crystals of \NaYbSe were synthesized by using a NaCl-flux method as reported previously \cite{Ranjith_NaYbSe2}.
We performed the \Na and \Se NMR measurements using a phase-coherent spectrometer, equipped with a 16 T superconducting magnet, and different cryostats operating in a variable temperature range between 0.3 and 300 K. We recorded the field-sweep NMR spectra by integrating spin-echo signals at fixed frequency, whereas at low temperatures, we recorded the broad-band spectra as sums of frequency-swept, Fourier-transformed spectra. 

We measured the nuclear spin-lattice relaxation rate \SL using a standard inversion-recovery technique, where the nuclear magnetization $m(t)$ was obtained from the recovery of the integrated spin-echo magnitude by variation of the delay time $\tau$ between the initial inversion pulse ($\pi$ pulse) and the $\pi/2 - \pi$ spin-echo sequence. We determined $T_1$ by describing the data with a stretched exponential function $m(t) = m(\infty) [1 - A (0.1 e^{-(t/T_1)^{\beta}} + 0.9 e^{-(6t/T_1)^{\beta}})]$ for the central transition ($I_z = -1/2 \leftrightarrow +1/2$), where $\beta$ is a stretching exponent, $m(\infty)$ is the nuclear magnetization at equilibrium ($t \rightarrow \infty$), and $A$ is the inversion factor. 

In order to obtain a sufficient sample mass, we stacked several single crystals. We used a silver wire for the sample coil to avoid an extrinsic \Cu NMR signal. We calibrated the external magnetic field by using the \Al NMR signal from a small piece of thin aluminum foil.
The in-plane and out-of-plane field orientations of the sample were set by using a single-axis goniometer, adjusting a minimum or maximum of the anisotropic resonance frequency or resonance field of the \Na satellite transitions, respectively.

\section{Results and discussion}

\begin{figure}[!ht]
\centering
\includegraphics[clip,width=0.8\columnwidth]{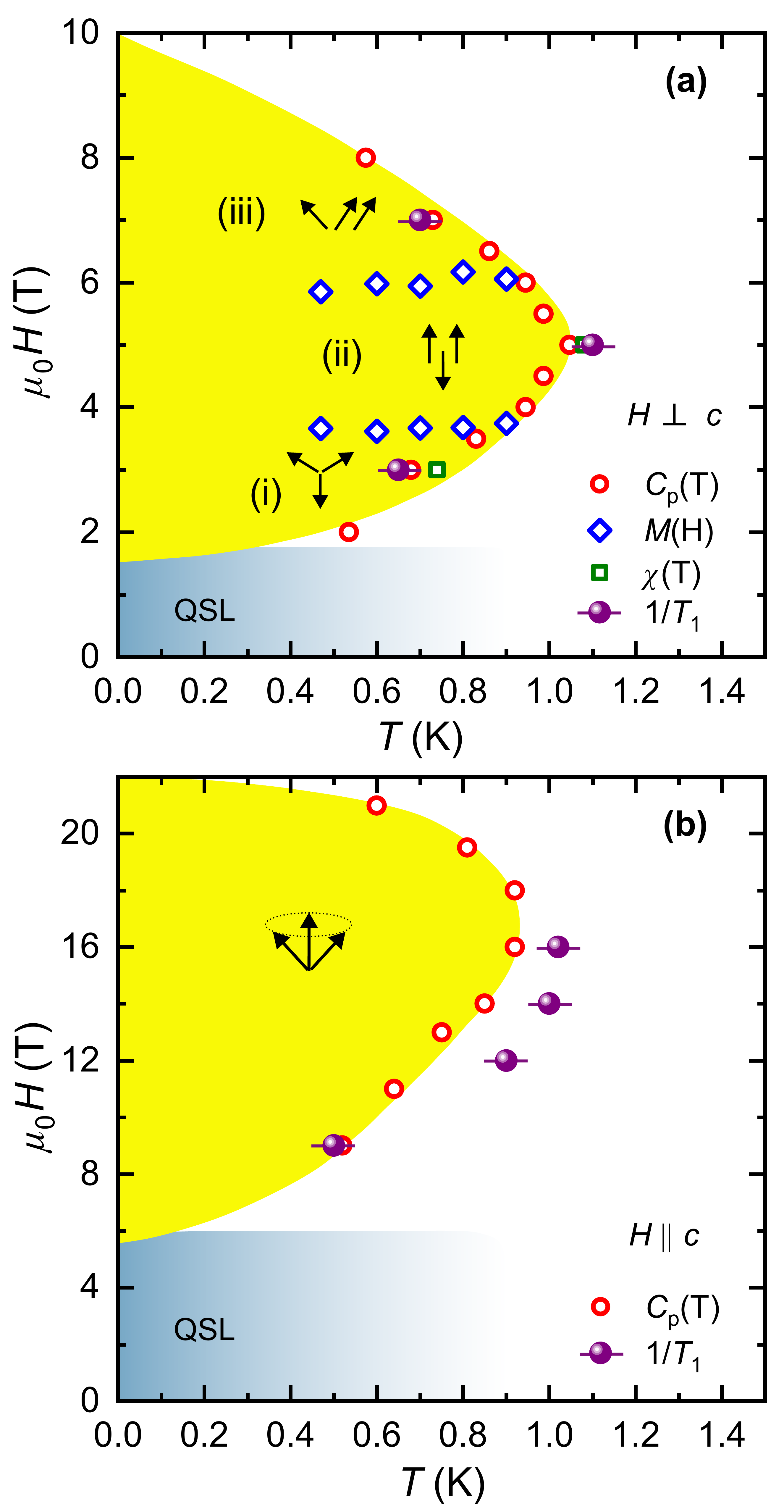}
\caption{Field-temperature phase diagram of \NaYbSe for (a) \Hperpc and (b) \Hparc, adapted from Ref. \cite{Ranjith_NaYbSe2}. The open symbols represent the phase transitions determined by means of different thermodynamic methods. The closed symbols represent the maximum of the \Na \SL rate. The phase transition temperatures obtained from specific heat and NMR differ by up to about 0.1 K, which corresponds to the temperature step size of the NMR. The blue low-field area represents the quantum spin liquid (QSL) region, while the yellow area represents the magnetically ordered phase with different spin configurations.}
\label{fig:phase_diagram}
\end{figure}

In Figs. \ref{fig:phase_diagram}(a) and \ref{fig:phase_diagram}(b), we show the field-temperature phase diagram of \NaYbSe for \Hperpc and \Hparc, respectively \cite{Ranjith_NaYbSe2}.
Here, the QSL regime at low fields is indicated by a blue fading area. The QSL is characterized by an increase in spin fluctuations, as probed by the \SL rate, which is discussed in the following sections.  At elevated fields, an ordered magnetic phase with different spin configurations is observed, as was previously reported \cite{Ranjith_NaYbSe2}.
The phase boundaries were determined by means of different thermodynamic methods.
As a new result of the present work, we show the phase transitions determined by means of the \Na nuclear spin-lattice relaxation rate.
The NMR results provide a microscopic confirmation of the field-induced magnetic ordered phases for both directions of applied magnetic field.

\begin{figure}[!ht]
\centering
\includegraphics[clip,width=0.75\columnwidth]{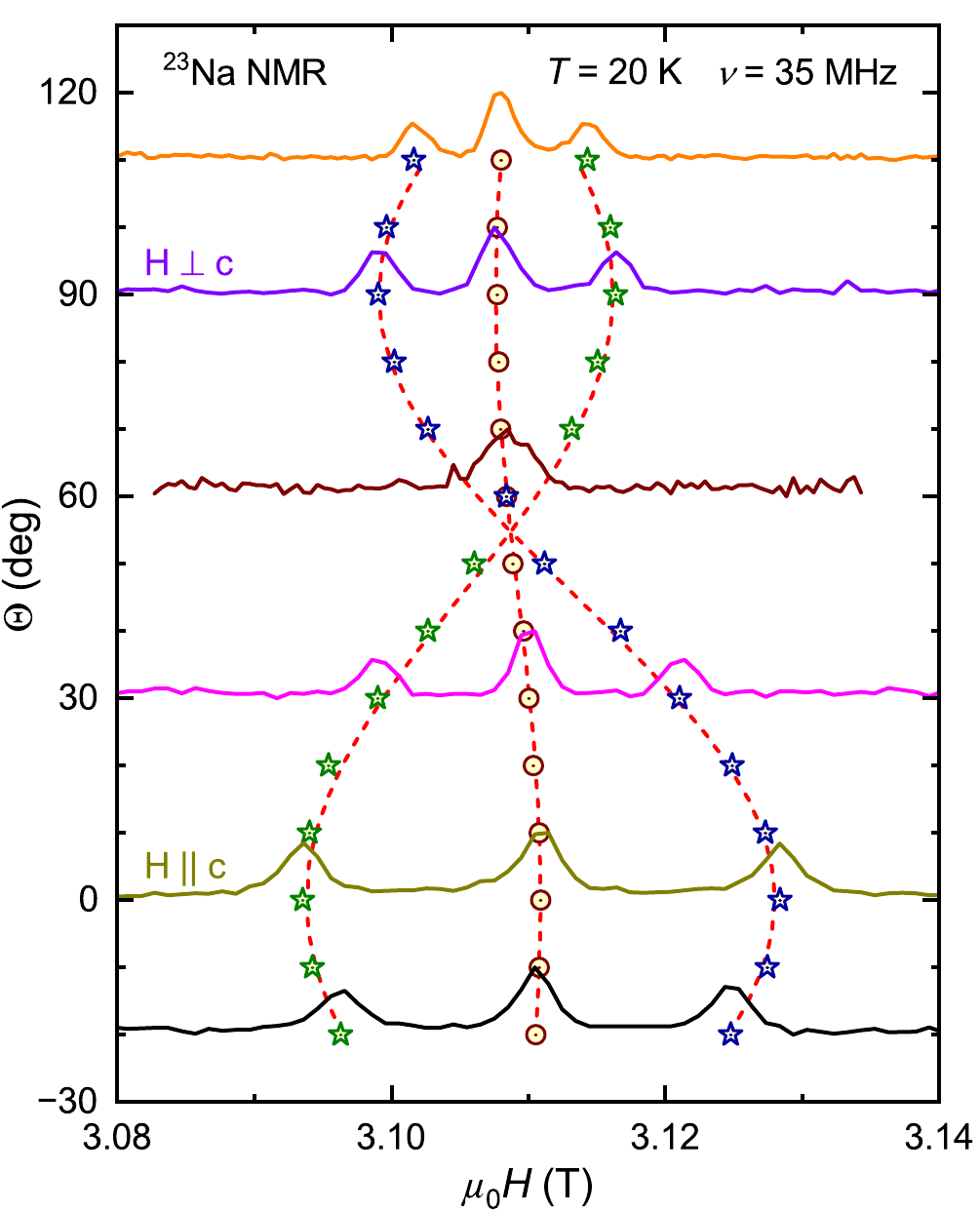}
\caption{\Na-NMR spectra for different field orientations including in-plane and out-of-plane direction, recorded at 20 K and 35 MHz. For the sake of clarity, only selected spectra are shown, whereas the symbols indicate the three transitions per \Na NMR spectrum for all measured spectra. The red dashed lines follow the angle-dependent resonance fields according to Eq. (\ref{equ:theta}).}
\label{fig:Na_angle}
\end{figure}

\subsection{\NaT NMR in the paramagnetic regime}

The isotope \Na has a nuclear angular moment of $I = 3/2$. Thus, we expect three spectral lines, a central line ($I_z = -1/2 \leftrightarrow +1/2$ transition)
and two satellites ($-3/2 \leftrightarrow -1/2$ and $+3/2 \leftrightarrow +1/2$ transitions), which are non-degenerate for an interaction of the electric nuclear quadrupole moment with a local crystal-electric field gradient. We can determine the magnetic and quadrupolar contributions to the resonance shift from orientation-dependent NMR measurements of a single-crystalline sample in an external magnetic field. We measured the angular-dependent \Na NMR spectra for a rotation between in-plane and out-of-plane orientations by varying the external magnetic field strength at $\nu = 35$ MHz and
20 K (Fig. \ref{fig:Na_angle}). here, in order to determine the magnetic and quadrupolar contribtutions to the resonance fields, we plot the spectral mass center as a function of angle $\Theta$ relative to the $c$ axis 
($\Theta = 0$).
We can write the contributions to the resonance fields $B_{\text{res}}(\Theta)$ of the satellites as

\begin{equation}
    B_{\text{res}}(\Theta) = B_{\text{m}}(\Theta) \pm \frac{\nu_{\text{Q}}}{2 \gamma_{\text{n}}} [3 \cos^2 (\Theta) - 1],
    \label{equ:theta}
\end{equation}
where $ B_{\text{m}}(\Theta)$ represents the anisotropic magnetic contribution, $\nu_{\text{Q}} \approx 200$ kHz is the nuclear quadrupole frequency, $\gamma_{\text{n}} = 11.2625$ MHz/T denotes the nuclear gyromagnetic ratio of \Na, and $\Theta$ is the angle of the external field relative to 
the $c$ axis \cite{Slichter_NMR}. For the central resonance field, only the contribution of $B_{\text{m}}(\Theta)$ to the angular dependence is considered, as the second-order quadrupole contribution can be neglected.

\begin{figure}
\centering
\includegraphics[clip,width=\columnwidth]{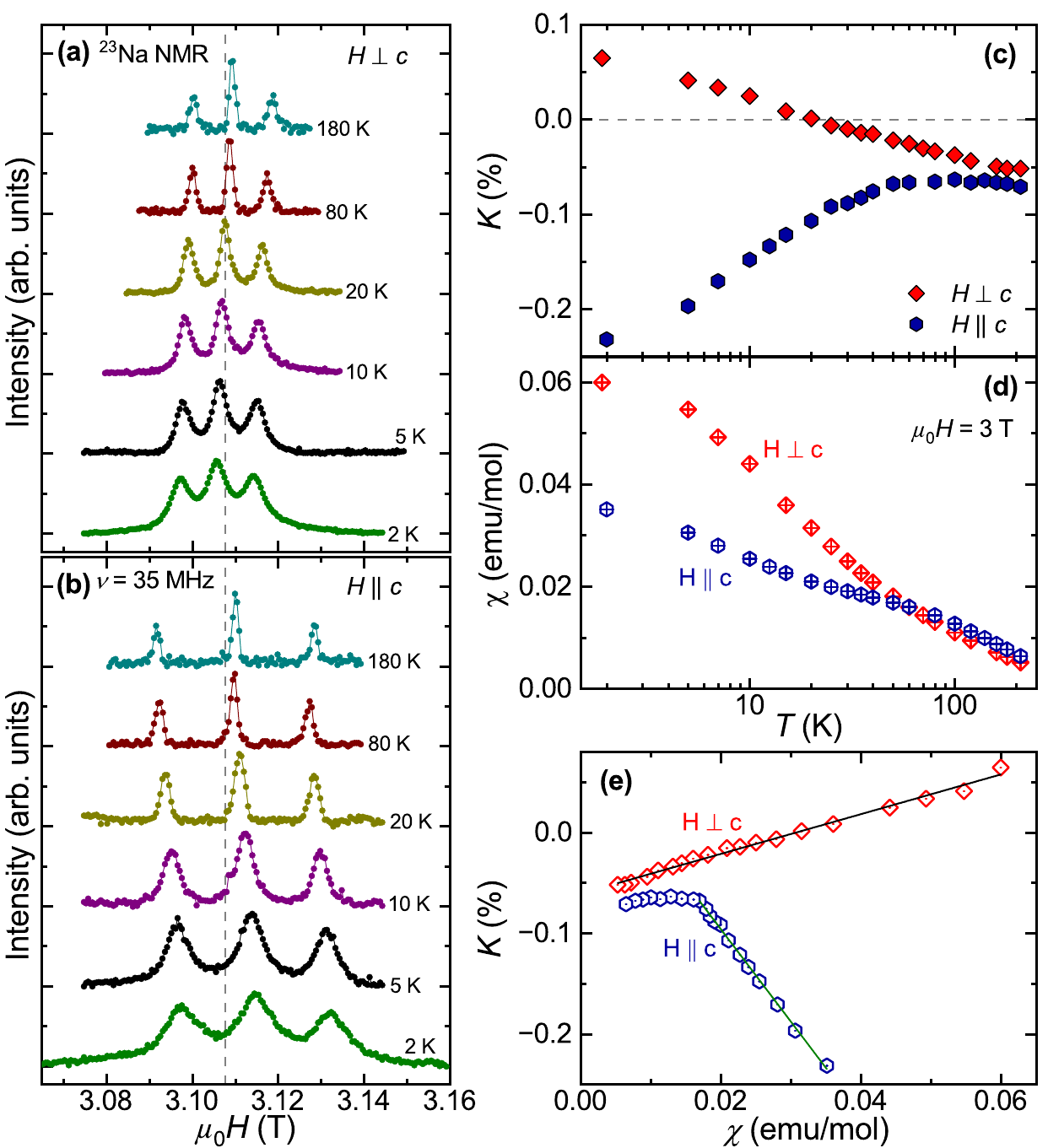}
\caption{Field-sweep \Na NMR spectra at selected temperatures in the paramagnetic regime for (a) \Hperpc and (b) \Hparc. The vertical dashed lines denote the reference field
$\mu_0 H_{\text{res}} = \nu / \gamma_{\text{n}}$. Temperature dependence of (c) Knight shift and (d) bulk dc susceptibility (3 T) for both field directions.
(e) Clogston-Jaccarino plot, constructed from the data in (c) and (d). Here, we determine the hyperfine coupling from a linear fit, as indicated by the solid lines.}
\label{fig:Na_hightemp}
\end{figure}

In Figs. \ref{fig:Na_hightemp}(a) and \ref{fig:Na_hightemp}(b), we present the \Na NMR spectra at selected temperatures for \Hperpc and \Hparc, respectively. 
The measurements were performed at $\nu = 35$ MHz by sweeping the external field around 3.1 T. 
With decreasing temperature, we observe a monotonous shift of the line positions and an increase of the spectral linewidth for both field directions. 
Figure \ref{fig:Na_hightemp}(c) shows the Knight shift $K$ for both field directions. Here, we observe the evolution of a pronounced anisotropy with decreasing temperature, where $K$
develops to positive values for \Hperpc and to negative values for \Hparc. We can determine the hyperfine constant $A_{\text{hf}}$ by constructing a Clogston-Jaccarino plot, where the Knight shift $K$ is plotted against the bulk susceptibility $\chi$ with the temperature as implicit parameter, yielding $K = A_{\text{hf}} \chi$ in the paramagnetic regime. 
We show the corresponding magnetic bulk susceptibility $\chi$ for both field directions in Fig. \ref{fig:Na_hightemp}(d), measured at a constant field of 3 T. Below 70 K, we observe an anisotropic behavior, as reported previously \cite{Ranjith_NaYbSe2}.

For analyzing the hyperfine couplings in a Clogston-Jaccarino-plot [Fig. \ref{fig:Na_hightemp}(e)], we observe the bulk dc susceptibility and the NMR Knight shift for both directions for an applied field of about 3 T. From a linear fit in the paramagnetic, high-temperature regime [Fig. \ref{fig:Na_hightemp}(e)], we determined the hyperfine coupling constants as the linear slope. For \Hperpc, we resolve no significant change of the slope across the whole data set, indicating that the hyperfine constant for this field direction remains unchanged across the whole temperature range, yielding $A_{\text{hf},\perp} = (11.0 \pm 0.2)$ mT/\muB. For \Hparc, we observe a different behavior for the high- and 
low-temperature regime. Here, we find a slope change at $K = -0.067$ and $\chi = 0.017$ emu/mol, corresponding to a temperature of 50 K. This indicates different hyperfine couplings for the different temperature regimes in this field direction. We attribute this slope change to the freezing of Kramers doublets and the resulting formation of a magnetic anisotropy at lower temperatures. Thus, we performed the linear fit only for temperatures below 50 K ($\chi > 0.017$ emu/mol) and yields $A_{\text{hf},||} = (-51.3 \pm 0.6)$ mT/\muB.

\begin{figure}
\centering
\includegraphics[clip,width=0.72\columnwidth]{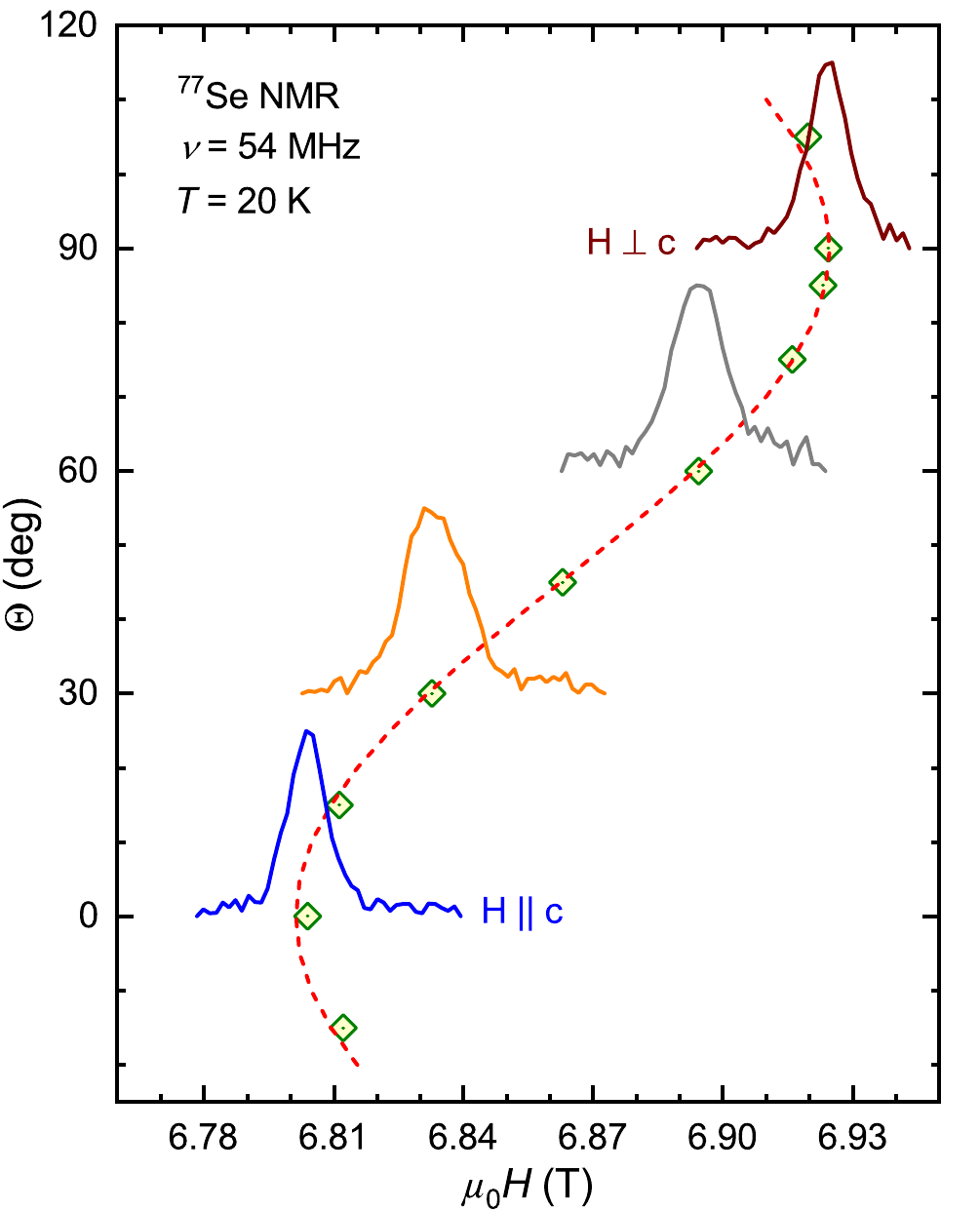}
\caption{\Se NMR spectra for different field orientations including in-plane and out-of-plane direction, measured at 20 K and $\nu$ = 54 MHz. Only selected \Se NMR spectra are shown, whereas the diamonds indicate the angular-dependent line positions of the \Se NMR for all measured spectra. The red dashed line indicates the
angular-dependent resonance field.}
\label{fig:Se_angle}
\end{figure}

\subsection{\SeT NMR in the paramagnetic regime}

The \Se isotope has an angular moment of $I = 1/2$, thus, we expect only one spectral line. In Fig. \ref{fig:Se_angle}, we present the angular-dependent \Se NMR spectra at selected angles, measured by sweeping the external magnetic field at a fixed frequency of $\nu = 54$ MHz and a temperature of 20 K. Since only one spectral line is observed, all \Se sites appear to be symmetrically equivalent in the crystallographic lattice. The angular dependence of the \Se resonance is given by the anisotropy of local magnetic fields, resulting in an angular-dependent shift of the spectral line between \Hparc and \Hperpc. We show this behavior in Fig. \ref{fig:Se_angle}, where we plot the corresponding resonance field for this spectral line as a function of angle $\Theta$ from the $c$ axis ($\Theta = 0$).

\begin{figure}
\centering
\includegraphics[clip,width=\columnwidth]{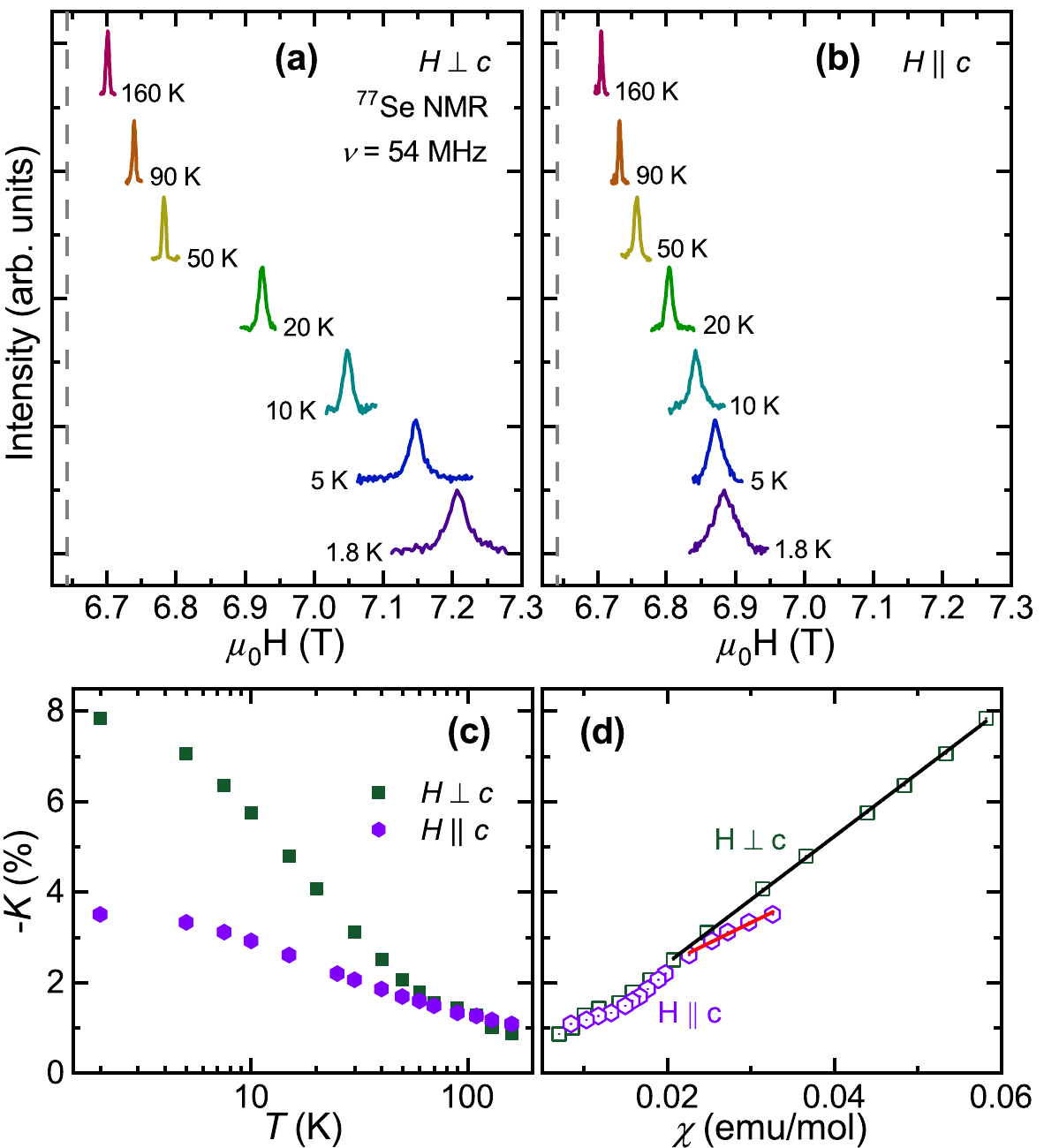}
\caption{Field-sweep \Se NMR spectra at selected temperatures for (a) \Hperpc and (b) \Hparc. The vertical dashed lines denote the reference field
$\mu_0 \text{H}_{res} = \nu / ^{77}\gamma$ at 54 MHz. (c) Temperature-dependent Knight shift for both field directions. (d) Clogston-Jaccarino plot for both field
directions. The solid lines are linear fits for the determination of the hyperfine couplings.}
\label{fig:Se_hightemp}
\end{figure}

In Figs. \ref{fig:Se_hightemp}(a) and \ref{fig:Se_hightemp}(b), we present the \Se NMR spectra at selected temperatures for \Hperpc and \Hparc, respectively. Towards low temperatures, we observe a monotonic shift of the spectral line position, accompanied by
an increase in linewidth. Below 60 K, a pronounced anisotropy of the Knight shift $K$ appears [Fig. \ref{fig:Se_hightemp}(c)], resembling our findings for the \Na shift. We determined the \Se hyperfine coupling constant as the slope of a linear fit in a Clogston-Jaccarino plot. For $\chi > 0.02$ emu/mol, corresponding to the low-temperature regime ($T \leq 30$ K) of the \Se shift, a change of the linear slope and thus, a change of the hyperfine coupling is evident for both directions [Fig. \ref{fig:Se_hightemp}(d)].
Again, we can attribute this behavior to the freezing of Kramers doublets at low temperatures. The hyperfine coupling constants are $A_{\text{hf},\perp} = (-782 \pm 7)$ mT/\muB for the in-plane direction
and $A_{\text{hf},||} = (-500 \pm 40)$ mT/\muB for the out-of-plane direction, respectively. The hyperfine coupling of \Se is stronger 
than that of \Na, likely due to the smaller distance between the electronic Yb moments and the nuclear \Se moments. The dominant mechanism of the hyperfine coupling is likely given by the superexchange between the Yb ions, mediated via the Se 4$p$ states. Notably, the anisotropy of the hyperfine field at the \Se sites follows that of the Yb ESR, with the largest fields perpendicular to the plane \cite{Sichelschmidt_gFactor_delafossite}, whereas it is the other way around for the \Na NMR results.

\subsection{QSL regime and field-induced order}

\begin{figure}
\centering
\includegraphics[clip,width=0.85\columnwidth]{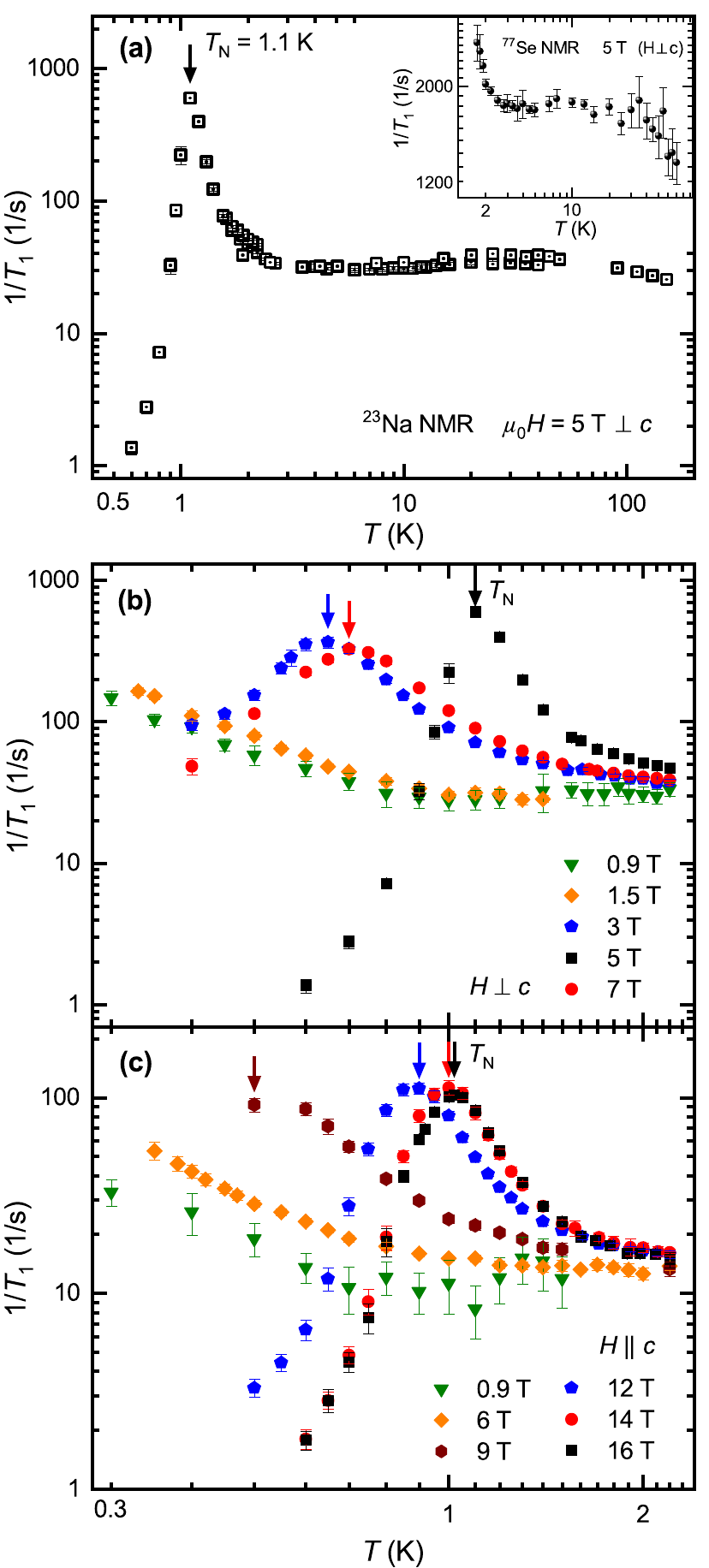}
\caption{(a) Temperature dependence of \Na \SL at 5 T (\Hperpc). Inset: \Se \SL at 5 T (\Hperpc) and temperatures above \TN. \Na \SL at low temperatures for (b) \Hperpc and 
(c) \Hparc at different magnetic fields. The arrows mark the transition temperature \TN into the magnetically ordered state, determined as maximum of \SL.}
\label{fig:Na_T1}
\end{figure}

We probed the low-energy spin fluctuations in \NaYbSeT by measuring the nuclear spin-lattice relaxation rate \SL, providing a sensitive and intrinsic means for the determination of transitions to the magnetically ordered phase. Figure \ref{fig:Na_T1} shows the temperature dependence of the \Na \SL, recorded for different field orientations. For an out-of-plane field of 5 T [Fig. \ref{fig:Na_T1}(a)], we observe a very shallow maximum between about 60 and 20 K, resulting from the freezing of Kramers doublets at
elevated energy levels. At lower temperatures, \SL decreases slightly down to 3 K. At temperatures below 3 K, \SL strongly increases, peaking at \TN = 1.1 K, which clearly signals the onset of long-range magnetic order. Below \TN, the opening of a spin excitation gap, arising from the magnetic order, results in a sharp drop of \SL.

In the inset of Fig. \ref{fig:Na_T1}(a), we show our results of the \Se \SL measurements at 5 T, performed between 80 and 1.5 K. Between 40 and 3 K, \SL is nearly temperature independent, whereas above 40 K, we observe a slight decrease
of the relaxation rate. Below 3 K, an increase of \SL is evident while approaching \TN. Compared to the \SL rate for \Na, we find a higher
relaxation rate due to the stronger coupling between the \Se nuclear moments and the Yb moments. At lower temperatures, the \Se NMR signal is wiped out due to a decrease of the $T_2$ decoherence time below the dead time of the spectrometer.

We present the data of the temperature-dependent \SL for the low-temperature regime and fields applied along \Hperpc and \Hparc in Figs. \ref{fig:Na_T1}(b) and \ref{fig:Na_T1}(c), 
respectively. In the low-field regime at 0.9 and 1.5 T, we observe a monotonous increase of \SL below 1 K, without
indications of magnetic order down to 0.35 K. We interpret this increase of the relaxation rate as critical spin fluctuations, arising from the QSL ground state in close vicinity of a magnetic instability. Upon application of higher magnetic fields, pronounced maxima of \SL are observed, which are clear signatures of transitions into the magnetically ordered phase. These results are in good agreement with the phase diagram determined by means of specific-heat measurements (Fig. \ref{fig:phase_diagram}) \cite{Ranjith_NaYbSe2}. 

\begin{figure}
\centering
\includegraphics[clip,width=\columnwidth]{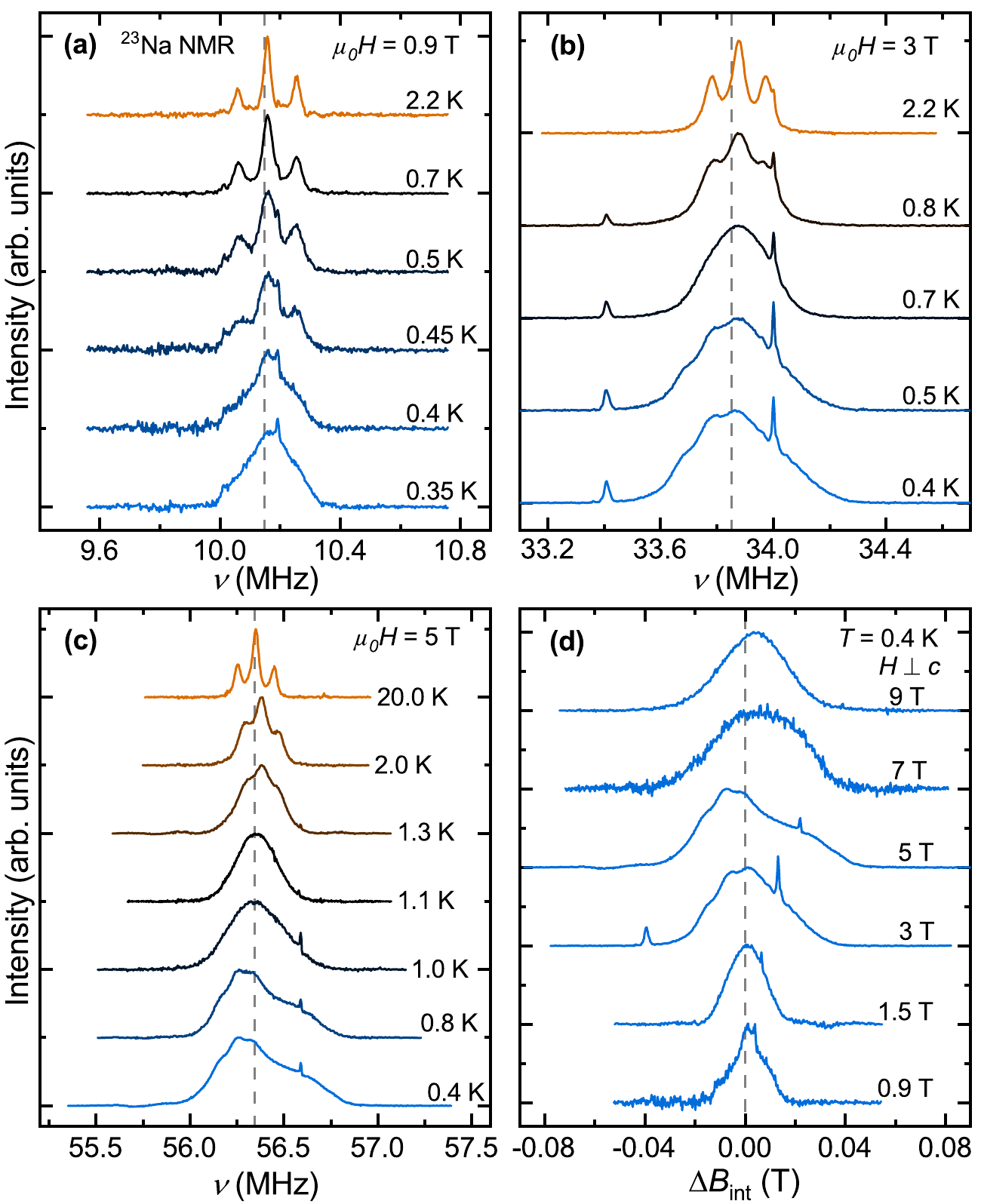}
\caption{Frequency-swept \Na spectra at selected low temperatures for \Hperpc at (a) 0.9 T, (b) 3 T, and (c) 5 T. The vertical dashed lines denote the \Na reference frequency. (d) Internal-field distribution $\Delta B$ of the \Na spectrum at 0.4 K (\Hperpc). Here, the reference was determined via a \Cu resonance.}
\label{fig:Na_lowtemp_ab}
\end{figure}

Figures \ref{fig:Na_lowtemp_ab}(a-c) depict the \Na spectra at low temperatures and different in-plane magnetic fields, recorded by sweeping the frequency at a constant magnetic field.
In Fig. \ref{fig:Na_lowtemp_ab}(a), we present the data for the lowest experimentally applied field of 0.9 T. Here, Ranjith \textit{et al.} suggested a putative QSL phase with increased spin fluctuations \cite{Ranjith_NaYbSe2}. At temperatures down to 0.7 K, the spectra exhibit three distinct lines without significant magnetic broadening. Below 0.5 K, we observe a homogeneous broadening of the spectral lines. This spectral broadening correlates with the increase in slow fluctuations, indicated by the enhanced \Na \SL rate [Fig. \ref{fig:Na_T1}(b)].

At 3 T, we observe a low-temperature increase of the \Na spectral linewidth, and the formation of an asymmetry in the high-frequency part of the spectrum [Fig. \ref{fig:Na_lowtemp_ab}(b)]. A 120$^{\circ}$ spin configuration was proposed for this field regime \cite{Ranjith_NaYbSe2}. The linewidth yields a significant increase below 1 K. Above 1 K, the \Na spectrum consists of three well-separated resonance lines, attributed to the quadrupolar interaction of the \Na nuclear moment with the local gradient of the crystal-electric field. Furthermore, a \Cu resonance appears in the spectra as an additional sharp line, serving as a reference for the \Na spectrum. Even though a silver wire was used for the
preparation of the RF coil, the appearance of a \Cu resonance may be due to the presence of copper-containing parts of the NMR probe near the RF coil.

For the \Na spectra at selected temperatures and 5 T [Fig. \ref{fig:Na_lowtemp_ab}(c)], we observe a symmetric increase of the linewidth below 4 K. Below 1 K, the spectral line exhibits clear asymmetric broadening, accompanied by a shift in the intensity maximum towards low frequencies. In this field region of the phase diagram [Fig. \ref{fig:phase_diagram}(a)], the magnetic order is characterized by an up-up-down spin configuration of the Yb moments \cite{Ranjith_NaYbSe2}. Such spin configurations, manifested as a plateau-like feature at 1/3 of the saturated bulk magnetization, are typical of frustrated spin-1/2 triangular-lattice systems \cite{Ba3CoSb2O9_magnetization_plateau_Shirata, 
Cs2CuBr4_magnetization_plateau_Ono, Xing_CsYbSe2, NaYbS2_magnetization_plateau}. 
Thus, we attribute the observed asymmetric broadening to the resulting local-field distribution within the crystal structure.

\begin{figure}
\centering
\includegraphics[clip,width=\columnwidth]{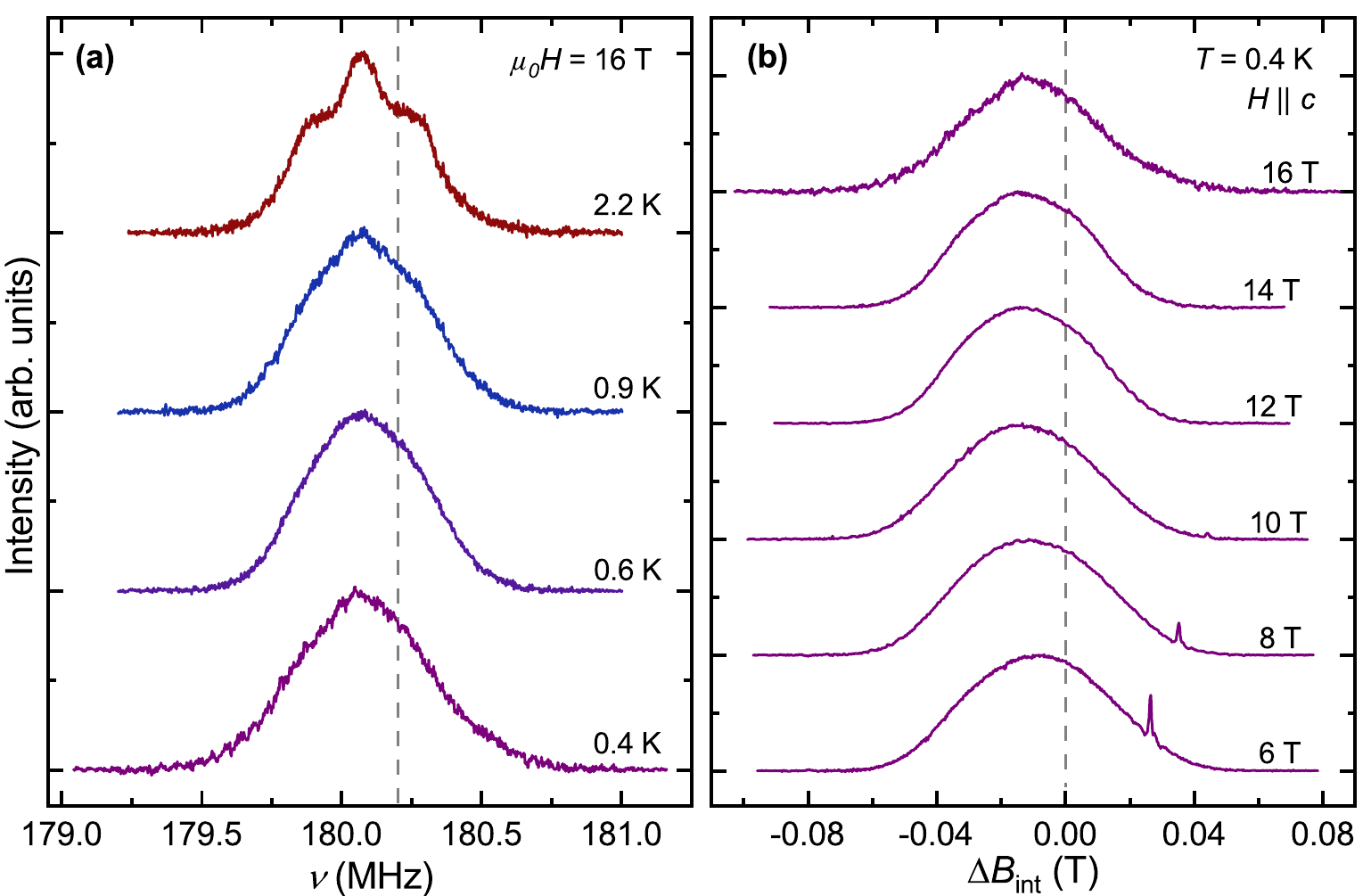}
\caption{(a) \Na low-temperature frequency-swept spectra at 16 T (\Hparc).  The vertical dashed lines denote the \Na reference frequency. (b) Internal-field distribution $\Delta B$ of the \Na spectrum at 0.4 K (\Hparc). Here, the reference was determined via the \Cu resonance.}
\label{fig:Na_lowtemp_c}
\end{figure}

In order to compare the internal-field distributions across the different regions of the phase diagram, we show the \Na spectra 
for increasing in-plane field amplitudes and at a constant temperature of 0.4 K in Fig. \ref{fig:Na_lowtemp_ab}(d). We determine the internal-field distribution at the \Na sites via the resonance frequency $\nu_0$, using $\Delta B_{\text{int}} = (\nu - \nu_0)/\gamma_{\text{n}}$, where $\nu_0 = 56.3125$ MHz and $\gamma_{\text{n}} = 11.2625$ MHz/T.
Thus, we can compare the field-driven evolution of the spectral shape, linewidth, and internal fields for the different parts of the phase diagram. Between 0.9 and 9 T, there is a systematic increase in spectral width. Within the field range of the QSL, at 0.9 and 1.5 T, the spectra exhibit a symmetric lineshape, 
indicating the absence of magnetic order. At 3 T, in the field regime of the 120$^{\circ}$ order, we observe a slight asymmetric broadening.
In contrast to the low-field regime, a strong increase in linewidth with an asymmetry towards higher frequencies indicates field-induced order. By further increase of the magnetic field to 5 T, a pronounced asymmetry in 
the spectral shape with a shift of the intensity maximum and the formation of a broad shoulder indicates a locally broken symmetry. At 7 T, where a canted up-up-down spin arrangement was proposed \cite{Ranjith_NaYbSe2}, a symmetrically broadened lineshape is evident, resembling the spectrum at 3 T. In the fully polarized regime at 9 T, we observe a symmetrically broadened spectrum.

Finally, we discuss our \Na NMR spectroscopy data for magnetic fields \Hparc. For this field 
orientation, specific-heat measurements indicated a field-induced ordered phase with a proposed umbrella-like spin configuration at elevated fields [Fig. \ref{fig:phase_diagram}(b)]. Figure \ref{fig:Na_lowtemp_c}(a) shows
the \Na spectra at selected temperatures and 16 T. The increased linewidth observed in the \Na spectra within the paramagnetic phase stems from the polarization of the electronic moments at elevated magnetic 
fields. Below 1 K, there is a symmetric spectral broadening as temperature decreases. In Fig. \ref{fig:Na_lowtemp_c}(b), we show the spectra at selected fields and constant temperature of 0.4 K for \Hparc. Across the entire range of the phase diagram from 
6 to 16 T, the spin configuration is of umbrella type, yielding the observed symmetric shape of the 
NMR spectra and an increased spectral linewidth. In contrast to the measurements along \Hperpc, we observe no asymmetric broadening of the \Na spectra.

\section{Conclusion}
Our comprehensive \Na and \Se NMR study covers the entire $H-T$ phase diagram of \NaYbSe. Our data provide strong indications for the presence of a QSL close to a magnetic instability in the low-field part of the phase diagram. Upon application of stronger magnetic fields, the NMR spin-lattice relaxation rate data prove the formation
of field-induced magnetic order. The asymmetric spectral broadening below the transition to the ordered state is consistent with the field-induced state and the spin textures postulated in this regime of the phase diagram. \NaYbSe is, thus, a prototypical system for studying the spin-liquid state on a triangular lattice with complex bond-dependent frustration. 
Perspectively, NMR studies in the millikelvin range, complemented by neutron scattering and muon spectroscopy, are essential for further advancing the understanding of the QSL state and the emergence of magnetic order. These investigations are crucial for establishing the hypothesized spinon liquid in this novel class of quantum magnets.

\section{Acknowledgments}
We thank P. Schlender for contributions in the beginning of the project and H. Yasuoka for discussions of the NMR results. Further thanks goes to H. Rosner and B. Schmidt for fruitful discussions on the theory of magnetic exchange in 4\textit{f} triangular lattices. We acknowledge the support from the Deutsche Forschungsgemeinschaft (DFG) through SFB 1143 (project ID 247310070) and the W\"{u}rzburg-Dresden Cluster of Excellence on Complexity 
and Topology in Quantum Matter - ct.qmat (EXC 2147, project ID 390858490). Further, we acknowledge the support of
the HLD at HZDR, member of the European Magnetic Field
Laboratory (EMFL).

\bibliography{literature}

\end{document}